# Surface Exciton Polaritons and Near-Zero Permittivity Surface Waves Supported by Artificial Organic Hyperbolic Metamaterials

*José N. Gama, Diogo Cunha, Carla Estévez-Varela, Marina García-Pardo, Pablo Pedreira, Adelaide Miranda, M. Carmen López-González, Pieter A.A. De Beule, Eduardo Solano, Rosalia Serna, Mikhail Vasilevskiy, Martin Lopez-Garcia, Isabel Pastoriza-Santos\*, Sara Núñez-Sánchez\**

José N. Gama, Isabel Pastoriza-Santos*
CINBIO, Universidade de Vigo, Campus Universitario As Lagoas, Marcosende, Vigo 36310, Spain
E-mail: pastoriza@uvigo.gal

Diogo Cunha, Mikhail Vasilevskiy
Centro de Física das Universidades do Minho e do Porto (CF-UM-UP), Universidade do Minho, Campus de Gualtar, Braga 4710-057, Portugal

Diogo Cunha, Adelaide Miranda, Pieter A.A. De Beule, Mikhail Vasilevskiy
International Iberian Nanotechnology Laboratory (INL), Avenida Mestre José Veiga s/n, Braga 4715-330, Portugal

Carla Estévez-Varela, Sara Núñez-Sánchez*
Instituto de Ciencia de Materiales de Madrid (ICMM), Consejo Superior de Investigaciones Científicas (CSIC), Sor Juana Ines de la Cruz, 3 Cantoblanco, Madrid 28049, Spain
E-mail: s.nunez.sanchez@csic.es

Marina García-Pardo, M. Carmen López-González, Rosalia Serna, Martin Lopez-Garcia
Instituto de Óptica (IO), Consejo Superior de Investigaciones Cientìficas (CSIC), Serrano 121, Madrid 28006, Spain

Pablo Pedreira, Eduardo Solano
ALBA Synchrotron Light Source, Cerdanyola del Vallès, Barcelona 08290, Spain

**Abstract text.** Hyperbolic metamaterials enable extreme light confinement and control of photonic states, but their realization has been restricted to inorganic architectures. Here, a fully organic route to fabricate artificial hyperbolic metamaterials based on multilayered thin films of J-aggregate carbocyanine dyes alternated with polyelectrolytes is introduced. These structures exhibit strong optical anisotropy and experimentally support hyperbolic surface exciton polaritons and, for selected dyes, additional surface waves in near-zero permittivity regimes. Spectroscopic ellipsometry confirms a uniaxial dielectric tensor with negative in-plane and positive out-of-plane components, close to the absorption peaks of the constituent J-aggregates. This anisotropy is preserved across individual layers, demonstrating the robustness of the layer-by-layer approach and enabling the coupling of surface exciton polaritons and near-zero permittivity modes even in films only a few nanometres thick. Transfer-matrix simulations based on the obtained dielectric tensor reproduce the coupling conditions for all thicknesses, validating the optical model. Structural characterization reveals the link between optical anisotropy and supramolecular order, with preferential in-plane molecular orientation and the evolution from discrete nanostructures to continuous films as deposition progresses. These organic hyperbolic metamaterial architectures, associated with narrow excitonic resonances from J-aggregates, offer a unique platform for tailoring emission, energy transport, and exploring polariton dynamics at the nanoscale.

## 1. Introduction

Artificial hyperbolic metamaterials (HMMs) are highly anisotropic media in which the real part of the in- and out-of-plane components of the dielectric tensor have opposite signs. This unique property leads to a hyperbolic dispersion relation, enabling propagation of high-wavevector modes and enhanced photonic density of states. Conventional plasmonic metamaterials are typically realized using metal-dielectric composites, such as multilayers or nanowire arrays, but their tunability is limited by the intrinsic optical properties of the constituent materials.[1] Nevertheless, HMMs are highly attractive for applications ranging from super-resolution imaging to emission control.[2,3] However, organic-based HMMs remain largely unexplored, despite their potential for molecular-level design and integration into flexible photonic platforms.

Organic semiconductors are a compelling alternative for HMMs, as they often naturally exhibit strong optical anisotropy and can support hyperbolic optical properties. The orientation of the transition dipole momenta within the molecular structure plays a critical role in determining their optical and electronic properties. Hence, the geometrical assembly of the constituent molecules directly impacts the device performance.[4,5] Several studies have demonstrated that spin-coated films of polymethine dyes, such as squaraine and quinoidal oligothiophenes, can display natural hyperbolic dispersion due to spontaneous lamellar ordering[6-8], with negative in-plane and positive out-of-plane components of the dielectric tensor. In these systems, anisotropy arises intrinsically from the spontaneous molecular packing.[9] For example, Barnes and coworkers[11] demonstrated that neat spin-coated films of a carbocyanine dye – TDBC – exhibit giant optical anisotropy and support hyperbolic surface exciton polaritons (h-SEPs), with distinct phase singularities observable via ellipsometric prism coupling.

However, the degree of anisotropy in spin-coated organic films is highly sensitive to monomer assembly dynamics on the substrate, often limiting control over molecular orientation during deposition.[10] Layer-by-Layer (LbL) assembly offers a robust alternative for the fabrication of artificial HMMs, enabling precise control over thickness and composition, and is especially suited for chromophores whose assembly can be controlled in solution.[12-15] J-aggregates are a notable example, their conformation can be controlled in solution and Frenkel exciton delocalization within each aggregate produces a narrow-band absorption spectrum.[16-18] This narrow-band absorption enables J-aggregate-based materials to achieve negative values of the real part of the dielectric tensor.[19] Therefore, combining J-aggregates with the alternating LbL deposition to construct a true multilayer architecture is especially promising for achieving hyperbolic dispersion, as this approach directly mimics the structural principles used in plasmonic HMMs and enables precise control over the anisotropic optical response.

In this work, we demonstrate the fabrication of fully organic artificial HMMs based on carbocyanine J-aggregates by optimizing an electrostatic LbL deposition protocol, as a scalable technique extendable to a virtual unlimited number of molecular families. This approach overcomes the limitations of uncontrolled molecular assembly such as spin-coating, obtaining uniaxial films with highly anisotropic optical properties. By controlling the supramolecular conformation of the dyes in solution, we achieved the deposition of J-aggregates with a preferential in-plane orientation of the transient dipole momenta. Our multiscale analysis, correlating molecular composition, nanoscale morphology, and molecular ordering confirms that the structural ordering induced by the LbL deposition is responsible for the hyperbolic response in the bulk, which is corroborated by both the anisotropic dielectric tensor obtained by ellipsometry and the excellent agreement between experimental high-momentum angular reflectivity and transfer-matrix simulated reflectance. The versatility of the method is demonstrated by obtaining artificial HMMs for three different dyes operating at specific wavelengths, with different hyperbolic dielectric tensors, where not only h-SEPs are supported, but also surface waves close to near-zero permittivity conditions.

## 2. Results and Discussion
### 2.1. Preparation and Characterization of Artificial Organic HMMs

To fabricate artificial organic HMMs, three dyes from a carbocyanine family with J-aggregate absorption bands at 561, 587, and 619 nm were selected, hereafter referred to as j560, j590 (also known in the scientific community as TDBC), and j620 (**Figure 1**b). These dyes carry sulfonate groups (Figure 1a), which provide them a net negative electrostatic charge and enable the assembly of J-aggregate multilayers via electrostatic LbL method. Using this approach, we alternated the deposition of the negatively charged J-aggregates with a positively charged polyelectrolyte, such as poly(diallyldimethylammonium chloride) (PDDA), as shown in Figure 1c, to construct highly ordered multilayered thin films on fused silica and borosilicate substrates. Each deposition cycle consisted of immersing the substrate in an aqueous PDDA solution followed by immersion in a J-aggregate dispersion. This cycle was repeated to achieve precise control over the film thickness, with each cycle referred to as one deposition. J-aggregate thin films with 1 to 20 depositions were analysed individually and a final PDDA layer was deposited to stabilise the last J-aggregate layer.

Transmission of thin films showed a primary minimum at wavelengths corresponding to the J-aggregate peaks in solution, which is deeper with the number of depositions (**Figures** 1d and **S1**). In the case of j560 thin films, they showed two distinct minima at 515 nm ($J^1_{560}$) and at 561 nm ($J^2_{560}$), which is consistent with the presence of two J-aggregate conformations at similar wavelengths that in their spectrum in solution. This analysis confirmed the successful transfer and stabilisation of the J-aggregate in the thin films in each deposition cycle for the three dyes.

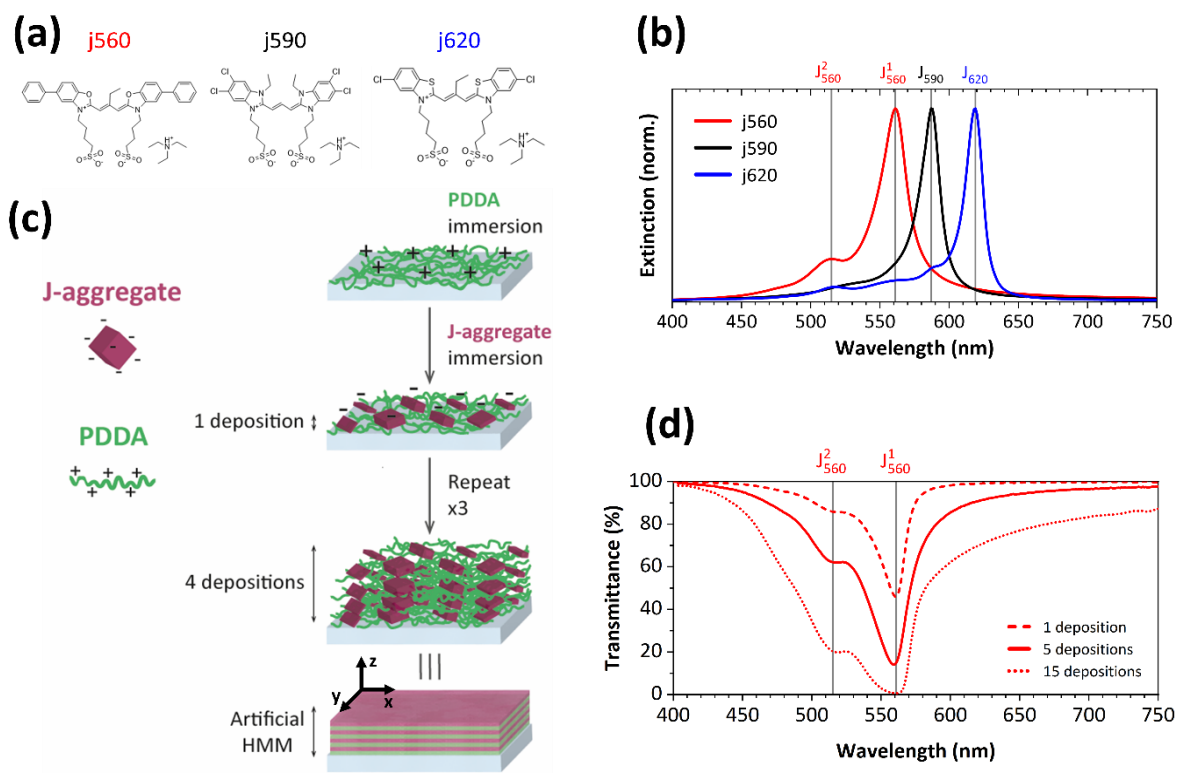

**Figure 1. (a)** Molecular structure of the selected carbocyanine dyes (j560, j590, and j620). **(b)** Normalised extinction spectra of the corresponding J-aggregate dyes in aqueous NaCl solutions. Vertical lines indicate the peak position of the J-aggregate bands: 515 nm ($J^2_{560}$) and 561 nm ($J^1_{560}$) for j560, 587 nm ($J_{590}$) for j590 and 619 nm ($J_{620}$) for j620 (section 4.3). **(c)** Scheme of the LbL assembly process for fabricating J-aggregate thin films (artificial HMMs) illustrating alternating deposition of the positively charged polyelectrolyte (PDDA) and negatively charged J-aggregates for controlled multilayer growth. **(d)** Normalized VIS normal incidence transmission spectra of J-aggregate thin films with different number of LbL depositions for j560. Vertical lines match the wavelengths of the peaks associated to the J-aggregate (indicated as $J^1_{560}$ and $J^2_{560}$) conformations of the dyes, targeted in Figure 1b.

The resulting J-aggregate thin films were analysed by Atomic Force Microscopy (AFM). Interestingly, AFM topographic 2D images revealed distinct morphological features for the different dyes after the first deposition (**Figure 2**a). For instance, j560 formed a continuous granular film (Figure 2a.i), j590 exhibited grain-like formations (Figure 2a.ii), and j620 assembled into ribbon-like structures (Figure 2a.iii). Moreover, the root mean square (RMS) roughness for thin films with 1 deposition cycle increased related to the bare borosilicate substrate, further confirming successful growth of J-aggregates at the surface after the first deposition (**Table S1**). For j590 and j620, initial films were discontinuous, however, gaps between structures progressively closed with successive depositions. For example, after 6 depositions, the surface became homogenous, forming a continuous film, where individual supramolecular assemblies were no longer distinguishable (Figures 2b.ii and 2b.iii). AFM analysis also enabled the determination of the film's thickness as a function of the number of depositions. Thin films were scratched in four different areas using a sharp tool and height profiles were measured and averaged. Since samples with more than 3 depositions presented a continuous surface (**Figure S2**), quantitative analysis starting from this point revealed near-linear thickness growth for the three J-aggregates (Figure 2c). The corresponding growth rates are summarized in Table S1.

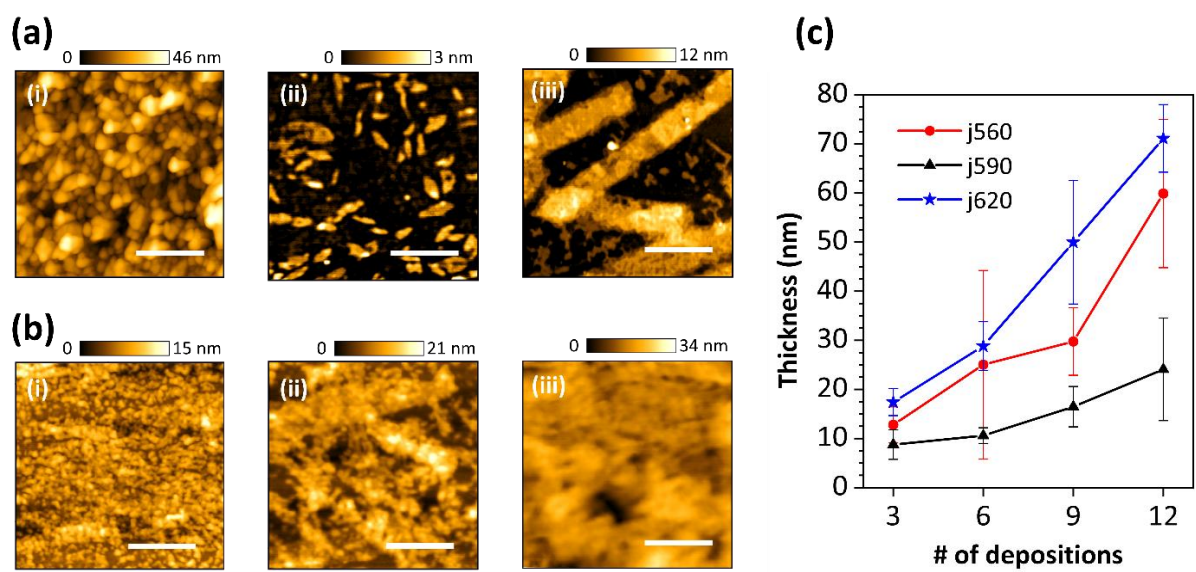

**Figure 2. (a-b)** AFM topographic 2D images of J-aggregate thin films after 1 **(a)** and 6 **(b)** depositions of j560 (i), j590 (ii) and j620 (iii). Scale bar: 1 μm. **(c)** Average film thickness as a function of the number of depositions for j560, j590, and j620. Error bars stand for 95% confidence standard deviation in a population of 4 measurements in different sample points.

To gain deeper insight into molecular packing and crystallinity, we performed synchrotron grazing-incidence wide-angle X-ray scattering (GIWAXS) analysis on j560, j590, and j620 thin films with increasing number of depositions (3, 5, 10, and 20). As shown in **Figure 3**a, the diffraction patterns of j560 display intense discrete peaks, growing in intensity with additional depositions, that is film thickness, primarily along the out-of-plane direction ($q_z$), parallel to the substrate. It is clearer when plotted the line profiles extracted along out-of- and in-plane directions, as shown in Figure 3b. This trend indicates enhanced and coherent lamellar stacking of monomers in the out-of-plane direction. Moreover, the characteristic real space distance (*d*) was calculated using $d = 2\pi/q$, where $q$ is the peak position in the reciprocal space. The main $q_z$ peak (Figure 3b, red line) corresponds to a lamellar spacing of 1.30 nm within the same ordered domain. Additionally, the presence of weaker peaks at 8 and 17 nm$^{-1}$ along the $q_r$ direction (Figure 3b, blue line) is consistent with molecular spacing governed by hydrogen bonding between monomers.[21] Ultimately, the absence of Debye-Scherrer rings, combined with distinct vertical and horizontal axes in the 2D patterns, provides strong evidence of a highly ordered and anisotropic material. GIWAXS thus suggests a uniaxial optical response, with J-aggregates randomly oriented in-plane but exhibiting preferential out-of-plane orientation. Consistent diffraction patterns recorded from 3 to 20 depositions confirmed that J-aggregate periodicity and orientation are preserved throughout film growth, with intensity increasing with thickness.

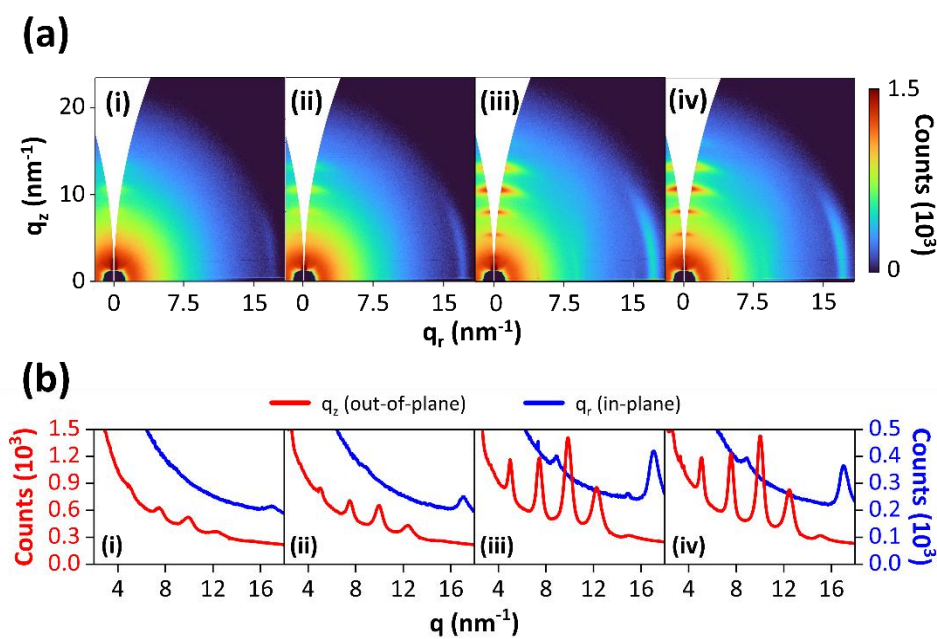

**Figure 3. (a)** Diffraction patterns obtained by GIWAXS of j560 thin films with (i) 3, (ii) 5, (iii) 10, and (iv) 20 depositions. Vertical and horizontal axes correspond to the in-plane ($q_r$) and out-of-plane ($q_z$) wavevector components, respectively. **(b)** Vertical and horizontal profiles extracted along the in- $q_z \approx 0$ nm$^{-1}$ (blue) and out-of-plane $q_r \approx 0$ nm$^{-1}$ (red) for the same samples, highlighting the evolution of diffraction peak intensity and position with increasing film thickness.

The same analysis on j590 and j620 films revealed similar trends (**Figures S3** and **S4**). In accordance with AFM measurements of discontinuous early depositions, diffraction peaks appeared only from 3 and 5 depositions, respectively. Both films exhibited the same uniaxial response and progressive intensity increase of diffraction peaks with additional depositions. These results demonstrate the robustness of the LbL deposition technique, as the persistence of identical interference fringes during growth indicates that ordered J-aggregate domains maintain tight spatial confinement within each layer, ultimately facilitating strong optical anisotropy in the films.

**2.2. Anisotropic Dielectric Tensor of Artificial Organic HMMs**

To further validate the structural anisotropy revealed by GIWAXS, variable-angle spectroscopic ellipsometry (VASE) was performed on the same set of J-aggregate thin films. VASE provides quantitative information on the complex dielectric tensors, which are directly influenced by molecular orientation and packing. Measurements were performed on J-aggregate thin films with different deposition cycles to ensure consistency of the determined complex dielectric tensor (section S2, Supporting Information, for more details). Importantly, the fitting procedure yielded consistent bulk dielectric tensor components across all thicknesses, with only the film thickness parameter adjusted according to AFM measurements. This demonstrates that the optical response of the films is thickness-independent within the analysed range, confirming the reliability of the LbL deposition method for producing homogeneous bulk properties.

The dielectric for the artificial HMMs of the three dyes was modelled as an anisotropic medium with a complex dielectric tensor. Following the structural analysis of the samples, we consider the artificial organic HMM as a highly uniaxial medium with an in-plane dielectric tensor $\varepsilon_{xx=yy}$ components and out-of-plane dielectric tensor component $\varepsilon_{zz}$ (**Figure S5**). The real and imaginary parts of the dielectric tensor components were modelled as a combination of Lorentz Oscillators, which together reproduce the frequency-dependent behaviour. The model parameters, oscillator amplitude, resonance energy, and damping, were optimised for each sample and all tensor components (**Table S2**), while the high-frequency dielectric constant ($\varepsilon_\infty$) was treated as a direct offset.

**Figure 4** shows the in-plane components of the dielectric tensor with respect to the optical axis of the thin film. As expected for a highly uniaxial material, the in-plane dielectric tensor $\varepsilon_{xx=yy}$ shows a strong absorption $\varepsilon''_{xx}>0$ with a region with $\varepsilon'_{xx}<0$, whereas the out-of-plane component $\varepsilon''_{zz}$ remains non-absorbing ($\varepsilon''_{zz} = 0$) for the j590 and j620, and only weakly absorbing for j560. The coloured areas in Figure 4a indicate the spectral regions where the dielectric tensor fulfils the hyperbolic condition ($\varepsilon_{xx}'\varepsilon_{zz}'< 0$) corresponding to Reststrahlen bands from 526 to 558 nm for j560 (Figure 4a.i), 529 to 583 nm for j590 (Figure 4a.ii), and 578 to 613 nm for j620 (Figure 4a.iii). Interestingly, for j560, a near-zero in-plane permittivity region component ($\varepsilon_{xx}' = 0$) appears between 500 to 528 nm, while the out-of-plane component remains positive.

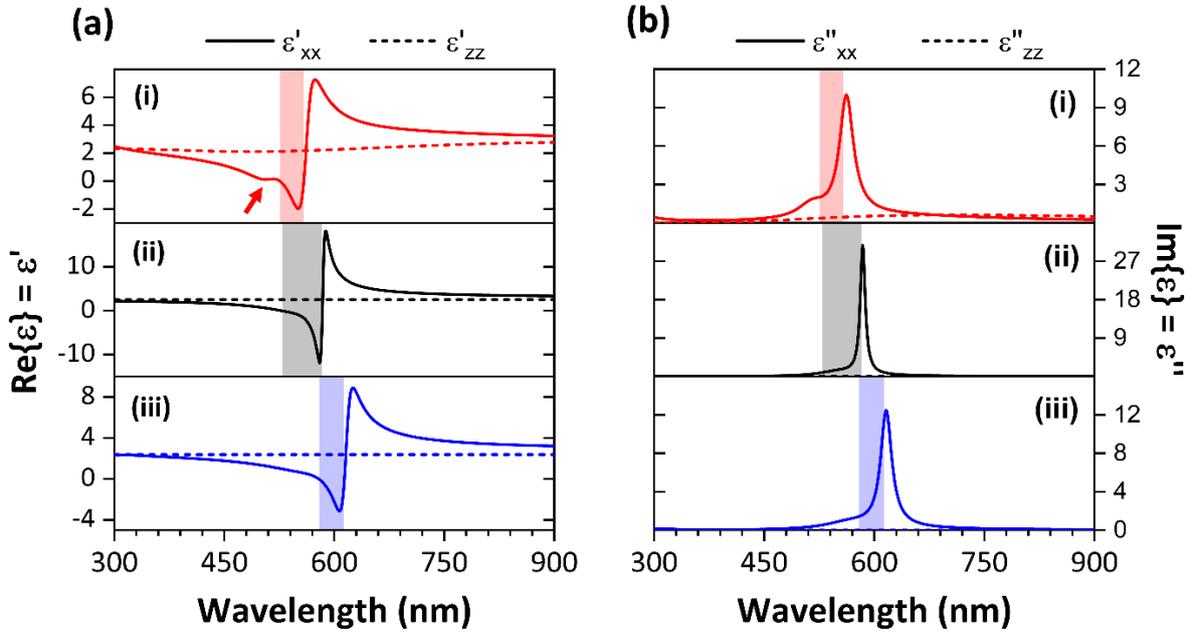

**Figure 4.** Fitted **(a)** real ε' and **(b)** imaginary ε'' parts of the dielectric tensor for the in-plane ($\varepsilon_{xx}$) and out-of-plane ($\varepsilon_{zz}$) components obtained from the ellipsometry analysis of J-aggregate/PDDA thin films of (i) j560, (ii) j590, and (iii) j620. Shaded regions indicate the Reststrahlen bands. The red arrow in **a.(i)** indicates the region with near-zero permittivity values for the case of j560.

## 2.3. Surface Exciton Polaritons and Near-Zero Permittivity Waves Supported by Artificial Organic HMMs

In this section, we measure the h-SEPs supported by the artificial organic HMMs and corroborated by theoretical simulations using the hyperbolic dielectric tensor obtained by ellipsometry in the previous section. **Figure 5**.a shows the reflectance spectra at different collected angles under a Kretschmann prism coupling configuration (see section 4.9 for more details) for J-aggregate thin films of the three dyes as a function of the number of depositions.[22,23] Above the prism critical angle ($\theta_c = 41.8°$), for j590 and j620, distinct single dips in the TM-polarized reflectance spectra are observed within the Reststrahlen bands of the dielectric's in-plane tensor. These dips are associated with the coupling conditions of the h-SEP mode, where the high-momentum photons are coupled to the guided mode at the film/air interface and not reflected. The dispersion of these features exhibited a clear dependence on film thickness (number of depositions). For j560 and j620 thin films with 1 deposition, a narrow coupling region is observed, indicating the formation of a well-defined h-SEP mode (Figures 5a.i and 5a.vii) in a film/nanostructure of just a few nanometres thick. In contrast, j590 required at least 2 depositions to support h-SEP modes (Figure 5a.iv).

As the number of depositions increases, the coupling dips become deeper and broader, and the mode profile evolves into a curved shape. Remarkably, for j560, a second resonance peak emerges at shorter wavelengths, becoming more pronounced with increasing thickness (Figure 5b.ii). This second resonance is a distinctive feature only of the j560-based artificial HMMs and occurs in the spectral region where the real part of the in-plane dielectric tensor approaches near-zero values. Such phenomenon, combined with strong anisotropy, can enable an extra surface mode. For reference, TE-polarized measurements are shown in **Figure S6**, revealing only guided mode coupling in TM-polarized conditions.

To further interpret the experimental data, we simulated the prism coupling conditions with a Transfer-Matrix Method (TMM) for anisotropic materials using the dielectric tensor obtained by ellipsometry for the films and tabulated borosilicate optical properties for the prism (see section 4.8 for more details). The film is considered a bulk anisotropic material with a hyperbolic dielectric tensor. The simulated TM-polarized reflectance spectra are in excellent agreement with the experimental results for any number of depositions (**Figure S7** for additional thicknesses), with reflectance dips becoming broader and more pronounced as the number of depositions increases (Figure 5.b). Notably for j560, the characteristic double-dip feature observed experimentally is well reproduced by the simulations. This demonstrates the robustness of the dielectric tensor of the artificial organic HMMs obtained by ellipsometry, which is consistent and independent of the number of depositions.

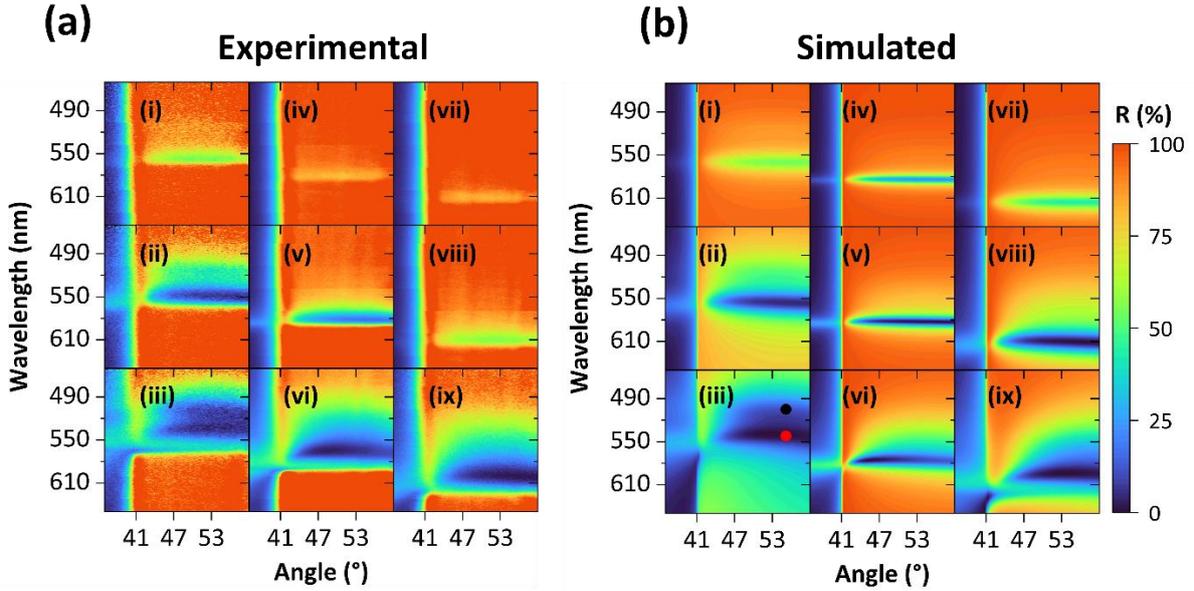

**Figure 5. (a)** Experimental and **(b)** simulated TM-polarized angular reflectance of J-aggregate thin films under surface waves coupling conditions for increasing deposition cycles. (i) 1, (ii) 5, and (iii) 15 j560 depositions; (iv) 2, (v) 5, (vi) 15 j590 depositions; (vii) 1, (viii) 5, and (ix) 15 j620 depositions. Theoretical modelling was performed using J-aggregate bulk film models, using the anisotropic permittivity obtained by ellipsometry and thicknesses estimated by AFM. A black and a red dot were placed at 505 and 540 nm, respectively, corresponding to the reflectance minima obtained for a j560 thin film with 15 depositions at an incident angle of 55°.

Electric field profile distributions were simulated at wavelengths and incidence angles corresponding to minima in TM-polarized reflectance under prism coupling conditions (Figure 5b). These simulations were performed to determine the nature of the coupled modes supported by the J-aggregate thin films and the surrounding media. For j590 and j620 films with 10 depositions (**Figure S8**), the electric field shows a maximum at the air/film interface and decays exponentially toward both the air and the thin film. This behaviour corroborates the presence of a strongly confined surface mode at the interface. The fact that this mode occurs at wavelengths where the real part of the in-plane dielectric tensor is negative confirms their attribution to h-SEPs.

In contrast, j560 thin films exhibited a distinctive double dip feature in the reflectance spectra, indicating coupling to two different modes. For a j560 thin film with 15 depositions (Figure 5b.iii) at an incidence angle of 55°, the first minimum appears at $\lambda = 505$ nm (black dot), and the second at $\lambda = 540$ nm (red dot). At 540 nm, the simulated electric field exhibits a pronounced evanescent decay inside the film and towards the surrounding media (**Figure 6**a), characteristic of a surface-bound mode [24]. This dip lies within the Reststrahlen band of the real part of the

j560 in-plane dielectric tensor and can therefore be attributed to the excitation of a h-SEP (red dot), as it happens for j590 and j620 films.

On the other hand, the first dip at 505 nm corresponds to the near-zero permittivity region, where the real part of the in-plane dielectric tensor approaches zero while its imaginary part remains significant, resulting in a partially confined polaritonic mode with extended field penetration (black dot). Thus, the field penetration is more extended, and the evanescent character is less pronounced (Figure 6b). Although the in-plane dielectric tensor is not negative at this wavelength, it approaches zero while its imaginary part remains significant, leading to a weakly confined, polariton-like mode supported by the lossy excitonic film. This interpretation is further supported by the calculated evanescent wavevector $\kappa_{2z}$ inside the film (**Table S3**), which shows that at 540 nm the evanescent contribution (real part of the complex momentum) dominates over the propagating one (imaginary part of the complex momentum), whereas at 505 nm both contributions are comparable. This dual behaviour with the coexistence of near-zero permittivity and hyperbolic regimes in j560 films reveals the versatility of organic artificial HMMs for controlling surface wave modes.

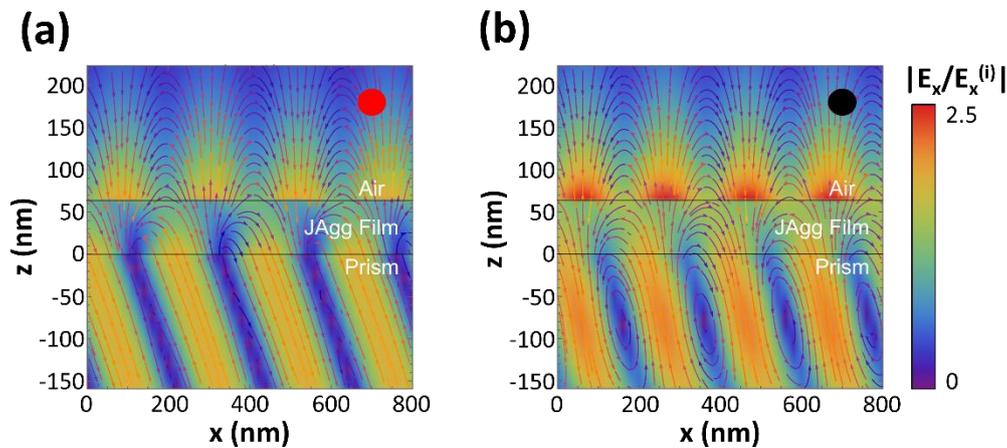

**Figure 6.** Two-dimensional colourmaps of the modulus of $E_x$ normalised by the incident electric field $E^{(i)}$ in a cross-section of a j560 thin film with 15 depositions with an incident angle of 55° for **(a)** $\lambda = 540$ nm (red line dot in Figure 5b.iii) and **(b)** $\lambda = 505$ nm (black dot in Figure 5b.iii). Displayed arrows correspond to the electric field vector $\vec{E} = (E_x, E_z)$.

## 3. Conclusions

In this work, we demonstrated a scalable LbL approach to produce artificial organic HMMs based on carbocyanine J-aggregates and polyelectrolyte linkers, achieving fine control over film thickness, composition, and molecular orientation. Through a comprehensive multiscale analysis, combining AFM, GIWAXS, and VASE, we confirmed the formation of highly ordered, anisotropic thin films with hyperbolic dielectric tensors preserved across individual layers, highlighting the robustness and reproducibility of the method.

Optical measurements revealed that all these films support h-SEP modes, with their spectral features dependent on the number of LbL depositions. For j590 and j620, the simulated modes are evanescent and well-confined at the film/air interface, characteristic of h-SEPs arising from a negative in-plane dielectric tensor. In contrast, j560 thin films exhibit a unique double-dip structure in the reflectance spectra, attributed to the coexistence of two distinct optical regimes: one in the near-zero permittivity region, supporting a weakly confined polariton-like mode, and another in the hyperbolic regime, supporting a confined h-SEP. Transfer-matrix simulations based on the experimentally determined permittivity reproduce these coupling conditions across thicknesses, validating the optical model.

Beyond structural and optical validation, these findings establish a scalable route to engineer light–matter interactions and excitonic transport in organic systems, enabling the design and control of distinct surface wave regimes, including hyperbolic exciton polaritons and near-zero permittivity surface modes. Theoretical simulations based on the transfer-matrix formalism and ellipsometric parameters show excellent agreement with experimental results. Overall, the strong correspondence between theory and experiment validates the reliability of our approach. Unlike conventional artificial HMMs typically based on inorganic metal-dielectric architectures, these organic systems synergize hyperbolic dispersion with strong excitonic resonances, offering a unique platform for ultrathin photonic devices, emission control, and exciton–photon coupling in flexible organic architectures.

## 4. Methods

### 4.1. Materials

Poly(diallyldimethylammonium chloride) (PDDA, Mw: 100,000-200,000, 20% wt. in $H_2O$) were purchased from Sigma Aldrich. Hydrogen peroxide ($H_2O_2$, 35% wt.) was purchased from Thermo Scientific. Sulfuric acid ($H_2SO_4$, 95% wt.) was purchased from Fisher Scientific. 5-Phenyl-2-[2-[[5-phenyl-3-(4-sulfobutyl)-3H-benzoxazol-2-ylidene]-methyl]-but-1-enyl]-3-(4-sulfobutyl)-benzoxazolium hydroxide, inner salt, sodium salt (j560), 5,6-dichloro-2-[[5,6-dichloro-1-ethyl-3-(4-sulfobutyl)-benzimidazol-2-ylidene]-propenyl]-1-ethyl-3-(4-

sulfobutyl)-benzimidazolium hydroxide sodium salt (j590), and 5-Chloro-2-[2-[5-chloro-3-(4-sulfobutyl)-3H-benzothiazol-2-ylidenemethyl]-but-1-enyl]-3-(4-sulfobutyl)-benzothiazol-3-ium hydroxide, inner salt, triethylammonium salt (j620) were provided by Few Chemicals GmbH. Absolute grade ethanol and milli-Q water (18.2 MΩ·cm) were used as solvents. 1 mm-thick microscope slides and thickness #1 coverslips were purchased from Marienfeld and epredia, respectively.

**4.2. Activation of Silica Substrate Surface**

Borosilicate and fused silica substrates were first washed with water and soap to remove grease, then treated with a piranha solution ($H_2SO_4/H_2O_2$, 4:1 v/v) for 20 minutes and rinsed twice in milli-Q water for 10 minutes each. This treatment eliminates organic contaminants and generates a high density of hydroxyl (-OH) groups on the surface, increasing surface energy and imparting a uniform negative charge.

**4.3. Control of Aggregation of Carbocyanine Dyes**

The preparation of stable J-aggregate solutions required careful and systematic optimization of aggregation conditions. For this work, we have selected three carbocyanine dyes, and we labelled them according to the centre wavelength of the corresponding J-aggregate: j560, j590, j620. For each selected dye, we established a rigorous three-step optimization protocol to achieve stable J-aggregate formation. Initial dissolution of the three dyes in ethanol (EtOH) distinct broadband extinction peaks for each (**Figure S9**). These spectra correspond to a solution of the dyes in monomer form. Subsequently, when the solvent was changed to water, a dramatic colour shift was immediately observed (**Figure S10**). UV-Vis spectroscopy revealed a red shift of more than 30 nm in the absorption peak compared to the bare monomer, accompanied by a sharpening of the band, which suggests the formation of ordered J-aggregate structures (Figures 1b and S10). This information was useful to label each dye as monomers according to the centre wavelength of the respective J-aggregate extinction maxima: j560, j590, j620. While j590 formed J-aggregates in milli-Q water without additional salts, the optimal aggregation for j560 and j620 required specific ionic strength conditions. A salt concentration screening with milli-Q water as solvent identified 50 mM for j560 (Figure S10a) and 100 mM for j620 (Figure S10b) as the ideal concentrations to maximize the J-aggregate absorption intensity while minimizing contributions from monomers and other colloidal conformations, such as H-aggregates.

### 4.4. Layer-by-Layer Deposition Protocol of Carbocyanine J-aggregates

As the activated substrate is negatively charged (see section 4.2), the first step is its immersion in an aqueous PDDA solution ([PDDA] = 1 mg/mL and [NaCl] = 0.5 M) for at least fifteen minutes. After this, the sample is washed thrice in milli-Q water (five seconds in each washing step) and dried with compressed air. The second step is to immerse the sample in a J-aggregate solution (milli-Q water as solvent, dye concentration of 300μM, and an ionic strength of 50, 0, and 100 mM for j560, j590, and j620, respectively (see section 4.3 for more details). The sample is kept under continuous agitation for five minutes to obtain a homogeneous deposition. After, the sample is dried with compressed air, before being washed once with a very quick immersion in an aqueous sodium chloride solution with the same concentration as in the optimised solutions for J-aggregation of each dye. Next, the sample is immersed with a very quick dip in the aqueous PDDA solution ([PDDA] = 1 mg/mL and [NaCl] = 0.5 M). At this point, we reach 1 deposition, and this cycle can be repeated n-times to achieve a thin film of n+1 depositions of J-Aggregate/PDDA. J-aggregate solutions must be prepared fresh from 5 to 5 depositions to ensure precise concentration of the dye.

### 4.5. Thin Film Characterization

UV-Vis absorption spectra were obtained using an Agilent 8453 UV-Vis spectrophotometer with glass and quartz cuvettes of 1 mm optical path. Transmission spectra were obtained using an Agilent Cary 5000. Atomic force microscopy (JPK Nanowizard 3 AFM, Bruker Nano GmbH, Germany) topographic measurements in quantitative imaging mode were obtained using a Standard Tapping mode PPP-NCHR (NanoSensors™) out-of-the-box cantilevers. The cantilevers with a nominal spring constant of 42 N/m have an aluminium coating layer on the detector side and a nominal frequency of 330 kHz. AFM images were acquired in Quantitative Imaging mode using a setpoint of 100-200 nN and a z-length of 60-300 nm. Image analysis was performed using Gwyddion 2.67.

### 4.6. Grazing-Incidence Wide-Angle X-Ray Scattering

GIWAXS measurements were performed at the ALBA synchrotron to investigate the crystallographic crystalline structure and molecular ordering of the J-aggregate thin films. The technique was selected for its well-known sensitivity to surface and near-surface structural features, which is particularly advantageous for characterizing very thin samples. For that, an X-ray beam of 12.4 keV was produced by an *in vacuo* undulator with 21mm period and 93 periods, and it was further monochromatized using a Si (111) channel cut monochromator. An array of Be compound refractive lenses (CRLs) was employed to collimate the beam, obtaining a beam size of 50 × 150 μm² (V × H) FWHM at the sample position. A stack of motors ensured precise tilts and translation of the sample in the beam. A Rayonix® LX255-HS detector with a pixel array of 2880 × 960 (V × H) pixels with a size of 88.54 × 88.54 μm² was employed to record the scattering patterns. The sample to detector distance and detector tilts were calibrated using $Cr_2O_3$ from NIST® as a calibration standard, allowing to generate the reciprocal space maps. To optimize the detection of diffraction peaks, especially in films with low deposition density, we systematically varied the grazing angle of incidence during the measurements. Only the optimized datasets, corresponding to angles that maximized the signal-to-noise ratio and peak visibility, are presented in this work: Figure 3, (i) 0.13°, (ii) 0.13°, (iii) 0.13°, and (iv) 0.11°; Figure S3 (i) 0.13°, (ii) 0.15°, (iii) 0.12°, and 0.13° (iv); and for Figure S4 (i) 0.13°, (ii) 0.13°, and (iii) 0.11°.

**4.7. Variable-Angle Spectroscopic Ellipsometry**

Spectroscopic ellipsometry characterization was performed on the J-aggregate-PDDA thin films deposited on fused silica substrates, in the 250-1700 nm wavelength range using a J. A. Woollam VASE system. The measurements of the ellipsometric parameters, Y and D, were done at three angles of incidence of 65º, 70º, 75º. Additionally, transmission (T) spectra at normal incidence (0º) were performed in the same film structure. We performed simultaneous fitting of the Y, D and T spectra with the transfer matrix formalism considering the films as an effective, homogeneous film on a semi-infinite fused-silica substrate. The dielectric tensor was modelled as uniaxial anisotropic to determine the in-plane ($\varepsilon_\parallel$) and out-of-plane ($\varepsilon_\perp$) components. The in-plane dielectric tensor ($\varepsilon_\parallel$) for each j-aggregate type was built as a sum of Lorentzian oscillators to ensure Kramers-Kronig consistent analysis. For the out-of-plane ($\varepsilon_\perp$) component for the films prepared with the j590, and j620 j-aggregates it has found that $\varepsilon_\perp$ = constant in all the studied range, whereas for the film prepared with j560 we found that it showed a weak and broad absorption that was well modelled with a Gaussian oscillator (section S2, Supporting Information). The best fit to the measured values was achieved with a Levenberg-Marquardt algorithm using the WVASE software. This has been proven to be a successful approach to determine the complex dielectric tensor when many fitting parameters are involved.[24]

## 4.8. Attenuated Total Reflectance and Electric Field Profile Calculations

The numerical calculations of the attenuated total reflection (ATR) spectra and electric field profiles shown were performed under TMM written in-house. Both reflectance maps and field profiles were derived from analytical equations described in section S3 of the Supporting Information, where more detailed information about the used models is written.

## 4.9. Surface Waves Coupling Measurements

The angle and polarization resolved reflectance of the J-aggregate LbL films was obtained using the Fourier Imaging Spectroscopy (FIS) technique. A halogen lamp covering the UV-Vis-NIR spectral range was focused down to a controllable spot size (~ 20 μm) by a high NA lens (Nikon 100x Plan Apo NA=1.45) that allowed the selection of single areas over the sample. The reflectance of the thin films was then collected by the same high NA lens in an epi-illumination configuration, and the back focal plane image of the lens was projected into a UV-Vis spectrograph (Princeton Instruments, Acton SpectraPro SP-2150 with attached CCD camera QImaging Retiga R6 USB3.0 Colour). SEP propagation can only be launch by matching the parallel momentum of the incident beam to that of the surface wave characteristic of an evanescent electric field at the J-aggregate/air interface. The high NA oil immersion lens (Nikon 100x Plan Apo NA = 1.45) allows both incidence and collection of reflected light with angles between -72.78° and 72.78°. Under this configuration, light was coupled to electromagnetic excitation in the thin film surface under Kretschmann prism-coupling equivalent configuration, which ensures coupling with wavevectors beyond the glass/air light line in k-space. The collected spectra were normalized by the spectra of a silver-coated mirror model -P01 >97% from THORLABS, Inc measured under the same conditions as the sample. The final normalized data were corrected by the reflectance spectra of the mirror calibrated by the producer at an angle of incidence (AOI) of 45º. The films were systematically prepared with increasing numbers of depositions on #1-thickness borosilicate coverslips (~ 0.17 mm-thick) for Surface Wave coupling measurements, considering the small working distance of the objective lens.

**Acknowledgements**

This work was supported by MCIN/AEI /10.13039/501100011033 and NextGenerationEU/ PRTR (Grants: TED2021-130522B-I00 and CNS2023-145364), the European Union under Grant Agreement #101129661-ADAPTATION and Xunta de Galicia/ERDF (Grant: GRC ED431C 2020/09). GIWAXS patterns were recorded at NCD-SWEET beamline with the collaboration of ALBA staff during the granted beamtimes 2022025594-2 & 2023097759. D.C. acknowledges FCT PhD Grant 2022.11947.BD (DOI: https://doi.org/10.54499/2022.11947.BD). S.N.-S. acknowledges the Severo Ochoa Centres of Excellence Program through Grant CEX2024-001445-S.

# Supporting Information

**Surface Exciton Polaritons and Near-Zero Permittivity Surface Waves Supported by Artificial Organic Hyperbolic Metamaterials**

*José N. Gama, Diogo Cunha, Carla Estévez-Varela, Marina García-Pardo, Pablo Pedreira, Adelaide Miranda, M. Carmen López-González, Pieter A.A. De Beule, Eduardo Solano, Rosalia Serna, Mikhail Vasilevskiy, Martin Lopez-Garcia, Isabel Pastoriza-Santos\*, Sara Núñez-Sánchez\**

## Section S1. Preparation and Structural Analysis of Artificial Organic HMMs

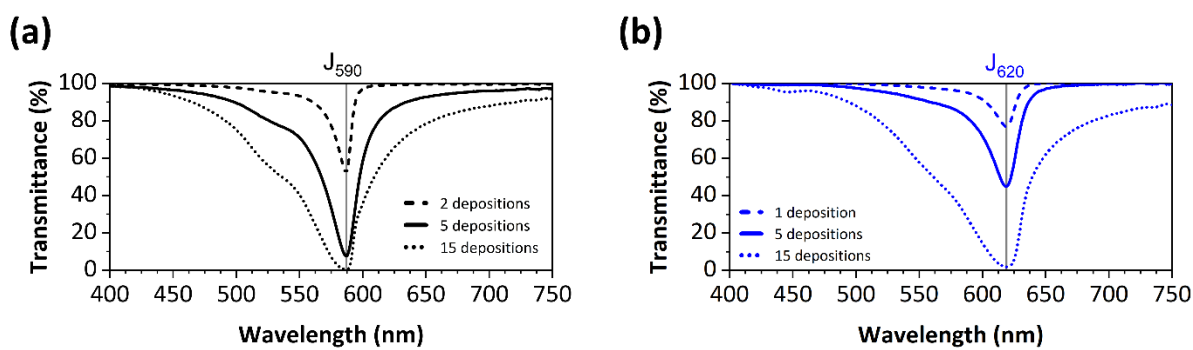

**Figure S1.** Normalized VIS normal incidence transmission spectra of J-aggregate thin films with different number of deposition cycles for: **(a)** j590 and **(b)** j620. Vertical lines match the wavelengths of the peaks associated to the J-aggregate (indicated as $J_{590}$ and $J_{620}$) conformation of the dyes, targeted in Figure 1b.

**Table S1.** Comparison of statistical data of thickness per deposition and RMS roughness of the first LbL deposition, obtained by analysis of AFM topographic 2D images of the thin films.

|  | **Borosilicate** | **j560** | **j590** | **j620** |
|---|---|---|---|---|
| **Thickness /dep. (nm)** | - | $4.25 \pm 1.49$ | $2.00 \pm 0.64$ | $5.58 \pm 0.91$ |
| **RMS roughness $r_q$ (nm)** | $0.35 \pm 0.06$ | $7.25 \pm 0.69$ | $0.46 \pm 0.18$ | $2.58 \pm 1.29$ |

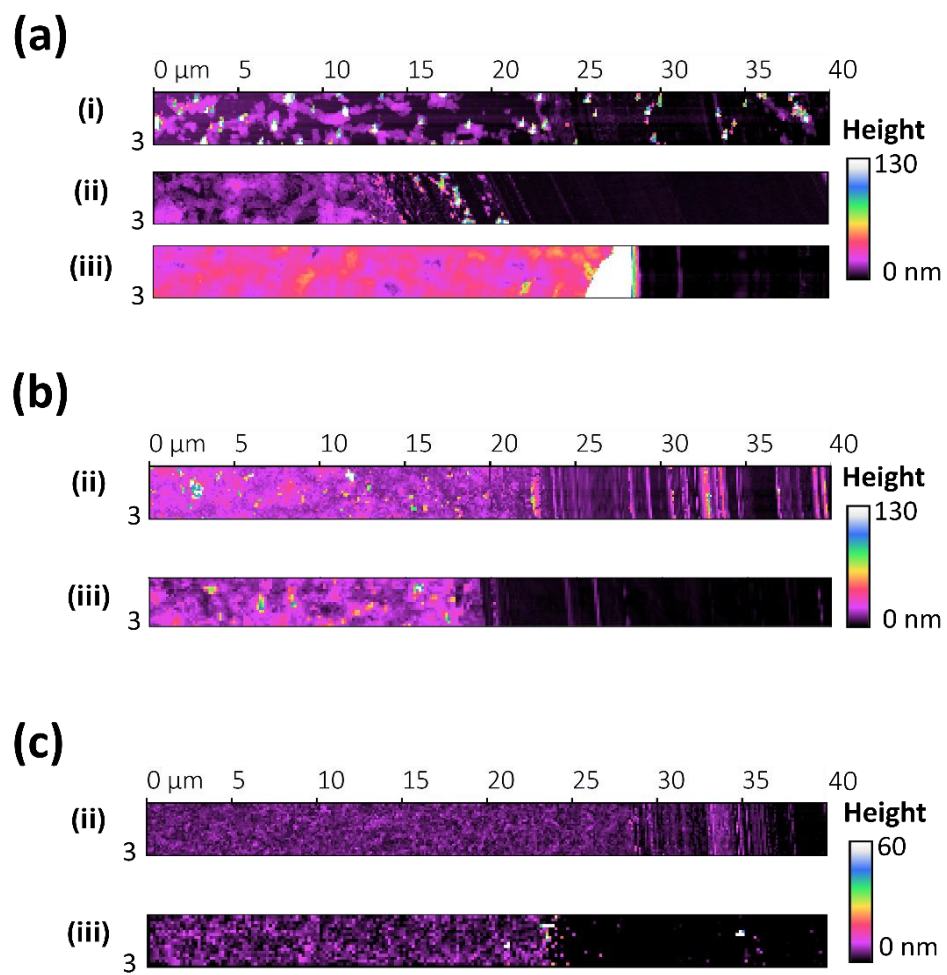

**Figure S2.** AFM height maps over scratched surface on thin films of **(a)** j560, **(b)** j590, and **(c)** j620 with: (i) 1, (ii) 3, and (iii) 6 depositions.

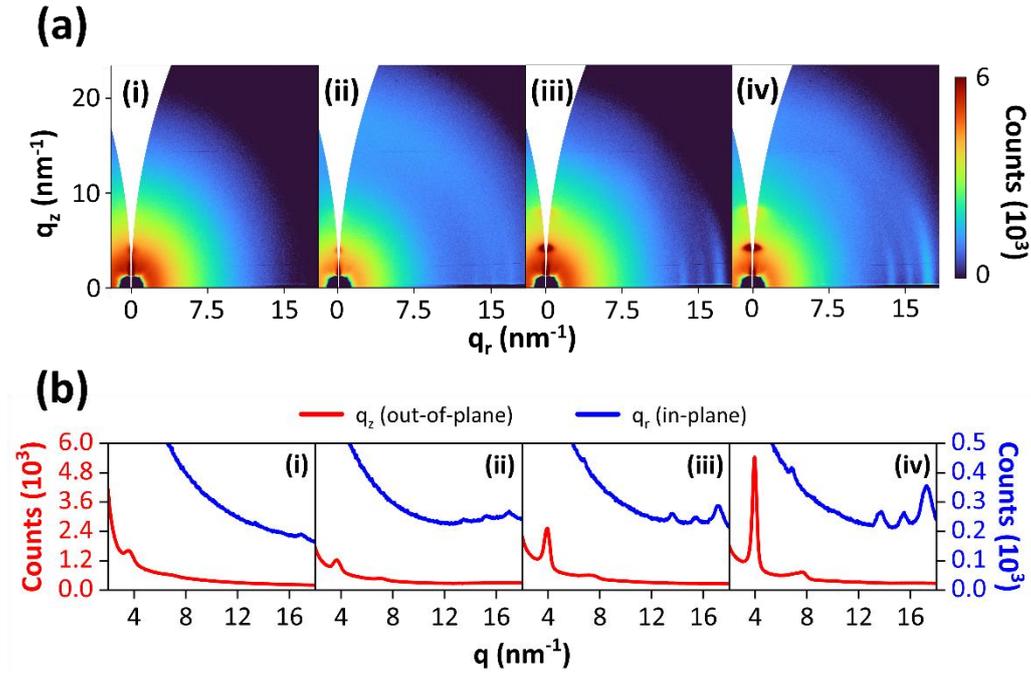

**Figure S3. (a)** Diffraction patterns obtained by GIWAXS of LbL thin films of j590 with (i) 3, (ii) 5, (iii) 10, and (iv) 20 depositions. Vertical and horizontal axes correspond to the in-plane and out-of-plane wavevector components, respectively. **(b)** Vertical and horizontal profiles taken from Figure 3a at near in-plane $q_z \approx 0$ nm$^{-1}$ (blue) and out-of-plane $q_r \approx 0$ nm$^{-1}$ (red) for (i) 3, (ii) 5, (iii) 10, and (iv) 20 depositions. The presence of a mild halo at $q_r = 16$ nm$^{-1}$ in figure S3a.(ii) is due to contributions of the substrate as the angle of incidence was tilted by 0.02° for better visualisation of the diffraction peaks.

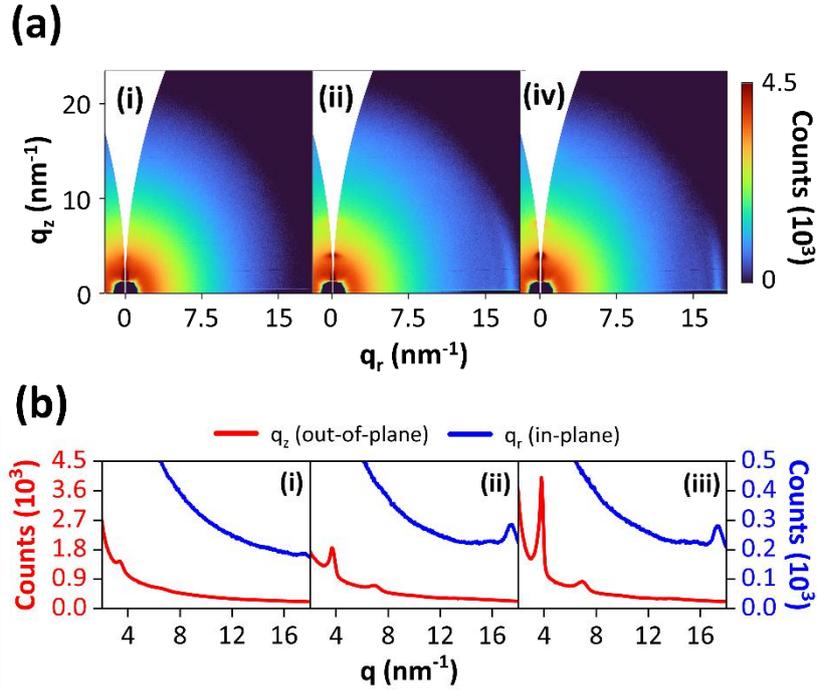

**Figure S4. (a)** Diffraction patterns obtained by GIWAXS of LbL thin films of j620 with (i) 5, (ii) 10, and (iii) 20 depositions. Vertical and horizontal axes correspond to the in-plane and out-of-plane wavevector components, respectively. **(b)** Vertical and horizontal profiles taken from Figure 3a at near in-plane $q_z \approx 0$ nm$^{-1}$ (blue) and out-of-plane $q_r \approx 0$ nm$^{-1}$ (red) for (i) 5, (ii) 10, and (iii) 20 depositions.

**Section S2. Optical response of Artificial Organic HMMs**

**Ellipsometry analysis**

The determination of the dielectric function of the J-aggregate/PDDA thin films was carried out by the analysis of VASE measurements. The ellipsometry measurements allow to obtain the $\Psi$ and $\Delta$ angles that are defined as:

$$\tan(\Psi)\, e^{i\Delta} = \frac{\tilde{r}_p}{\tilde{r}_s}$$

where $\tilde{r}_p$ and $\tilde{r}_s$ are the Fresnel coefficients for the parallel and perpedicular components respect to the plane of incidence. With the ellipsometer we measure $\tilde{r}_p$ and $\tilde{r}_s$ as a function of wavelength and angle of incidence ($\theta$).

Accurate fittings of the measured $\Psi$ and $\Delta$ values for the multilayered J-aggregate/PDDA thin film structures were obtained using the transfer matrix formalism and modelling the films with uniaxial anisotropy. In this model, the films are treated as uniaxial layers with the optical axis perpendicular to the substrate, as shown in Figure S5.

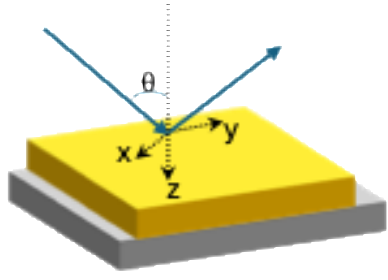

**Figure S5.** Scheme showing the axis defined for the uniaxial J-aggregate/PDDA thin film configuration. The optical axis is parallel to the z axis. The x and y axis define the film plane. The blue arrows show the incident light at an angle θ and define the plane of incidence.

According to this configuration, we define the in-plane dielectric tensor as $\varepsilon_{xx=yy}$ and the out-of-plane dielectric tensor as $\varepsilon_{zz}$. The corresponding dielectric tensor is therefore:

$$\boldsymbol{\varepsilon} = \begin{pmatrix} \varepsilon_{xx} & 0 & 0 \\ 0 & \varepsilon_{xx} & 0 \\ 0 & 0 & \varepsilon_{zz} \end{pmatrix}$$

To obtain the dielectric function for the J-aggregate/PDDA thin films of the three studied J-aggregate derivatives: j560, j590, and j620; we used a combination of oscillators which together reproduce the frequency-dependent behaviour of the complex dielectric function:

$$\varepsilon_n = \varepsilon'_n + i\varepsilon''_n$$

Each oscillator term contributes to the overall dielectric response and is expressed, in the Lorentzian form, as:

$$\varepsilon_n(\omega) = \frac{f_n \gamma_n \omega_n}{\omega_n^2 - \omega^2 - i\gamma_n \omega}$$

where $f_n$ is the oscillator strength, $\omega_n$ the resonance frequency, and $\gamma_n$ the damping constant. The final dielectric function is therefore obtained as a sum of the Lorentzian oscillators:

$$\varepsilon(\omega) = \varepsilon_\infty + \sum_n \varepsilon_n(\omega)$$

This approach ensured the full Kramers-Kronig consistency. Exceptionally, for example for the out-of-plane dielectric tensor of the j560 derivative, the experimental spectra exhibited broader absorbing features that were better described by a Gaussian oscillator. The fitting parameters and the corresponding oscillator types used for each film are listed in Table S10.

**Table S2.** Fitted parameters for the in-plane ($\varepsilon_{xx=yy}$) and out-of-plane ($\varepsilon_{zz}$) dielectric tensor components of j560, j590, and j620 thin films.

| | j560 | | | j590 | | | | j620 | | | |
|---|---|---|---|---|---|---|---|---|---|---|---|
| *In-plane component of dielectric tensor ($\varepsilon_{xx=yy}$)* | | | | | | | | | | | |
| Oscillator | $f_n$ | $\omega_n$ (eV) | $\gamma_n$ (eV) | Oscillator | $f_n$ | $\omega_n$ (eV) | $\gamma_n$ (eV) | Oscillator | $f_n$ | $\omega_n$ (eV) | $\gamma_n$ (eV) |
| Lorentz | 9.9 | 2.21 | 0.09 | Lorentz | 28.9 | 2.12 | 0.03 | Lorentz | 12.0 | 2.01 | 0.06 |
| Lorentz | 1.0 | 2.41 | 0.22 | Lorentz | 1.2 | 2.24 | 0.28 | Lorentz | 8.0 | 2.18 | 0.35 |
| Lorentz | 5.0 | 2.08 | 0.10 | Gauss | 9.1 | 6.33 | 0.46 | Lorentz | 0.9 | 4.03 | 0.55 |
| Lorentz | 2.3 | 4.83 | 0.63 | - | - | - | - | Gauss | 0.9 | 6.14 | 2.15 |
| *Out-of-plane component of dielectric tensor ($\varepsilon_{zz}$)* | | | | | | | | | | | |
| Oscillator | $f_n$ | $\omega_n$ (eV) | $\gamma_n$ (eV) | Oscillator | $f_n$ | $\omega_n$ (eV) | $\gamma_n$ (eV) | Oscillator | $f_n$ | $\omega_n$ (eV) | $\gamma_n$ (eV) |
| Gauss | 0.6 | 1.75 | 1.42 | No Oscillator | - | - | - | No Oscillator | - | - | - |

**Section S3. Transfer Matrix formalism for anisotropic media**

To compute the attenuated total reflectance (ATR) spectra for a setup closely resembling the experimental configuration in a consistent approach with the previous section, which is described in detail on section 4.9 of the main manuscript, we have modelled the J-aggregate/PDDA films as a uniaxial anisotropic layer, characterized by the dielectric tensor shown in section S2, where $\varepsilon_{xx=yy}$ and $\varepsilon_{zz}$ are the in-plane and out-of-plane dielectric tensors obtained from the ellipsometry analysis, respectively.

The scheme consists of a layer of oil serving as the prism, followed by a refractive index-matching glass coverslip. Beyond the coverslip lies the uniaxial J-aggregate film, and finally, a semi-infinite air medium. In this arrangement, the evanescent field generated at the prism interface couples to the excitonic transitions within the J-aggregate/PDDA thin film, leading to the formation of SEPs localized at the film-air interface. Since the refractive index of the immersion oil is chosen to match that of the glass substrate, this setup can be classified as a Kretschmann configuration. To compute the reflectance of the multilayer system, we employ the transfer matrix formalism, which relates the tangential components of the electromagnetic fields across each interface. For a uniaxial layer of thickness $d$, the transfer matrices written in the basis of the tangential field components ($H_y$, $E_x$), for TM and TE polarizations are given by:

$$T_{TM} = \begin{pmatrix} \cos(k_{2z}^{(e)} d) & i\frac{\omega \varepsilon_{xx}}{c k_{2z}^{(e)}} \sin(k_{2z}^{(e)} d) \\ i\frac{k_{2z}^{(e)} c}{\omega \varepsilon_{xx}} \sin(k_{2z}^{(e)} d) & \cos(k_{2z}^{(e)} d) \end{pmatrix}; T_{TE} = \begin{pmatrix} \cos(k_{2z}^{(o)} d) & i\frac{\omega}{c k_{2z}^{(o)}} \sin(k_{2z}^{(o)} d) \\ i\frac{k_{2z}^{(o)} c}{\omega} \sin(k_{2z}^{(o)} d) & \cos(k_{2z}^{(o)} d) \end{pmatrix}$$

where $k_{2z}^{(e)}$ and $k_{2z}^{(o)}$ are the longitudinal wavevector components for the extraordinary (TM) and ordinary (TE) modes in the uniaxial film, respectively. These are defined as:

$$k_{2,z}^{(e)^2} = \frac{\varepsilon_{xx}}{\varepsilon_{zz}}\left(\varepsilon_{zz}\frac{\omega^2}{c^2} - q^2\right); \quad k_{2,z}^{(o)^2} = \varepsilon_{xx}\frac{\omega^2}{c^2} - q^2$$

where the in-plane (transverse) wavevector is $q=\sqrt{\varepsilon_1}(\omega/c)\sin\theta_i$, with $\theta_i$ denoting the angle of incidence at the prism/J-aggregate interface. These matrices allow us to propagate the electromagnetic fields through the excitonic layer, from $z=0$ to $z=d$.

The total transfer matrix of the system can then be used to derive the transmission ($\hat{t}$) and reflection ($\hat{r}$) coefficients for each polarization. They are given by:

$$\hat{t}_{TM}=\frac{2}{T_{11}^{-1}+\frac{ck_{3z}}{\omega\varepsilon_3}T_{12}^{-1}+\frac{\omega\varepsilon_1}{ck_{1z}}T_{21}^{-1}+\frac{k_{3z}\varepsilon_1}{k_{1z}\varepsilon_3}T_{22}^{-1}}; \hat{t}_{TE}=\frac{2}{T_{11}^{-1}+\frac{ck_{3z}}{\omega}T_{12}^{-1}+\frac{\omega}{ck_{1z}}T_{21}^{-1}+\frac{k_{3z}}{k_{1z}}T_{22}^{-1}}$$

$$\hat{r}_{TM}=\left(T_{11}^{-1}+\frac{ck_{3z}}{\varepsilon_3\omega}T_{12}^{-1}\right)\hat{t}-1;\ \hat{r}_{TE}=\left(T_{11}^{-1}+\frac{ck_{3z}}{\omega}T_{12}^{-1}\right)\hat{t}-1$$

where media 1 and 3 correspond to the prism and air, respectively. The longitudinal wavevector components in these media are:

$$k_{1(3)z}=\sqrt{\varepsilon_{1(3)}\frac{\omega^2}{c^2}-q^2}$$

and $T_{ij}^{-1}$ denotes the $(i,j)$ element of the inverse transfer matrix associated with the corresponding polarization. Finally, the reflectance spectra for both polarizations are obtained as:

$$R_{TM}=|\hat{r}_{TM}|^2;\ R_{TE}=|\hat{r}_{TE}|^2$$

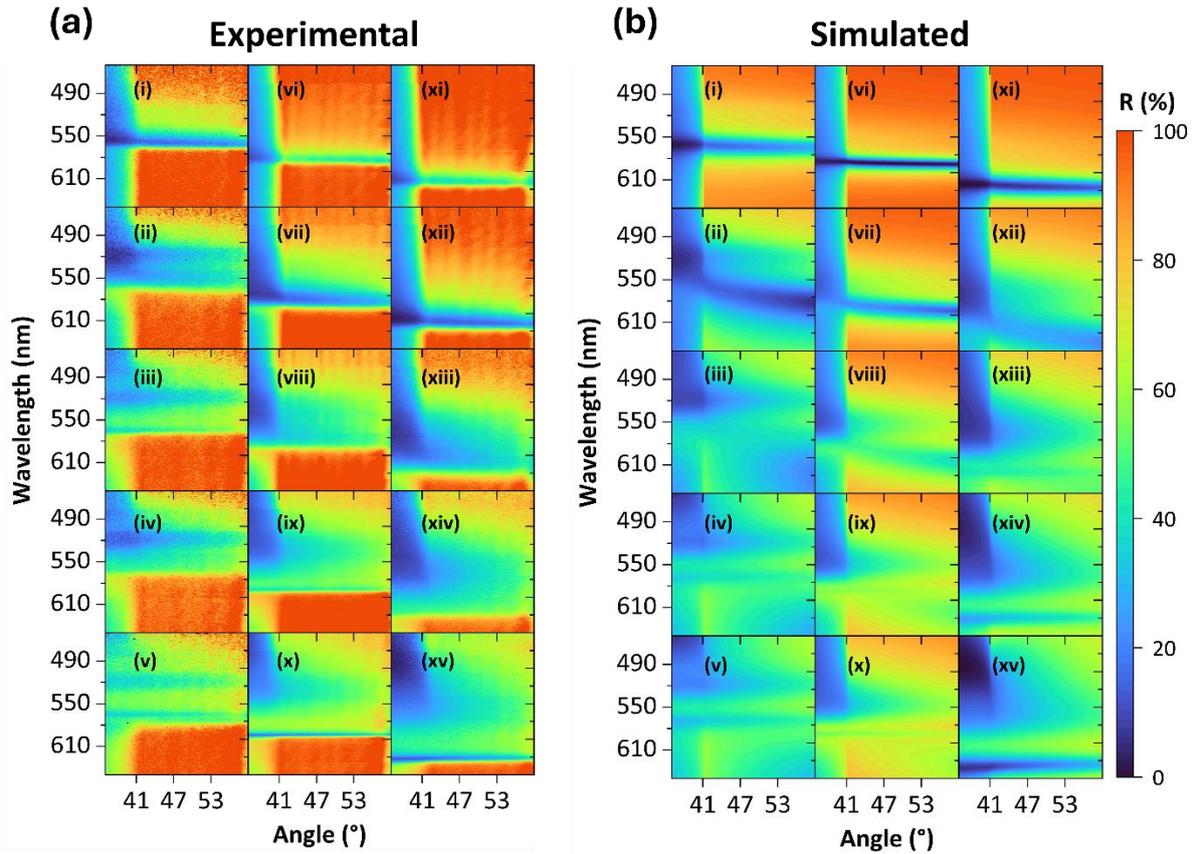

**Figure S6. (a)** Experimental and **(b)** simulated TE-polarized angular reflectance of J-aggregate thin films under surface waves coupling conditions for increasing deposition cycles. (i) 1, (ii) 5 (iii), (iv) 10, (v) 15, and (vi) 17 j560 depositions; (vii) 2, (v) 5, (viii) 10, (ix) 15, and (x) 20 j590 depositions; (xi) 1, (xii) 5, (xiii) 10, (xiv) 15, and (xv) 20 j620 depositions. Theoretical modelling was performed using J-aggregate bulk film models, using the anisotropic permittivity obtained by ellipsometry and thicknesses estimated by AFM.

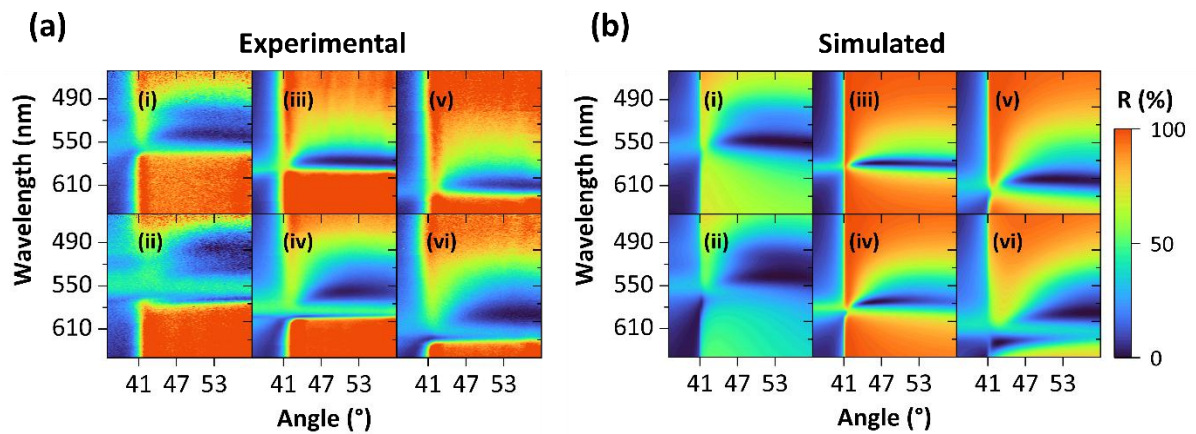

**Figure S7. (a)** Experimental and **(b)** simulated TM-polarized angular reflectance of J-aggregate thin films under surface waves coupling conditions for increasing deposition cycles. (i) 10 and (ii) 17 j560 depositions; (vii) 10 and (x) 20 j590 depositions; (xi) 10 and (xv) 20 j620

depositions. Theoretical modelling was performed using J-aggregate bulk film models, using the anisotropic permittivity obtained by ellipsometry and thicknesses estimated by AFM.

**Section S4. Electric Field Profiles**

To plot the field profile in the Kretschmann configuration, we first write the expressions for the electromagnetic fields in each region. Considering the experimental setup described in detail on section 4.9 of the main manuscript, in the oil-substrate region, located at $z<0$, the field is expressed as the sum of incident and reflected propagating waves:

$$\vec{E}_1 = \left[\left(E_{1,x}^{(i)}, 0, E_{1,z}^{(i)}\right)e^{ik_{1z}z} + \left(E_{1,x}^{(r)}, 0, E_{1,z}^{(r)}\right)e^{-ik_{1z}z}\right] e^{i(qx-\omega t)};$$
$$\vec{H}_1 = \left[\left(0, H_{1,y}^{(i)}, 0\right)e^{ik_{1z}z} + \left(0, H_{1,y}^{(r)}, 0\right)e^{-ik_{1z}z}\right] e^{i(qx-\omega t)}.$$

where $E^{(i)}$ and $E^{(r)}$ mean incident and reflected electric field. In the J-aggregate film, located at $0<z<d$, the field is a superposition of decaying and growing evanescent waves. Thus, we write:

$$\vec{E}_2 = \left[\left(E_{2,x}^{(+)}, 0, E_{2,z}^{(+)}\right)e^{\kappa_{2z}z} + \left(E_{2,x}^{(-)}, 0, E_{2,z}^{(-)}\right)e^{-\kappa_{2z}z}\right] e^{i(qx-\omega t)};$$
$$\vec{H}_2 = \left[\left(0, H_{2,y}^{(+)}, 0\right)e^{\kappa_{2z}z} + \left(0, H_{2,y}^{(-)}, 0\right)e^{-\kappa_{2z}z}\right] e^{i(qx-\omega t)}.$$

where $E^{(+)}$ and $E^{(-)}$ mean growing and evanescent waves. Finally, in the air region ($z>d$), the field is once again evanescent and can be written as:

$$\vec{E}_3 = \left(E_{3,x}, 0, E_{3,z}\right)e^{-\kappa_{3z}z} e^{i(qx-\omega t)};$$
$$\vec{H}_3 = \left(0, H_{3,y}, 0\right)e^{-\kappa_{3z}z} e^{i(qx-\omega t)}.$$

The wavevectors of the previous expressions are defined as:

$$k_{1z} = \sqrt{\varepsilon_1 \frac{\omega^2}{c^2} - q^2}; \quad \kappa_{2z} = \sqrt{\frac{\varepsilon_{xx}}{\varepsilon_{zz}}\left(q^2 - \varepsilon_{zz}\frac{\omega^2}{c^2}\right)}; \quad \kappa_{3z} = \sqrt{q^2 - \varepsilon_3 \frac{\omega^2}{c^2}}.$$

To compute the field profiles, we need first to relate the several field amplitudes. To do that we can apply boundary conditions. At $z=0$ we obtain:

$$E_{1x}^{(i)} + E_{1x}^{(r)} = E_{2x}^{(+)} + E_{2x}^{(-)};$$
$$H_{1y}^{(i)} + H_{1y}^{(r)} = H_{2y}^{(+)} + H_{2y}^{(-)},$$

and at $z=d$ we get:

$$E_{2x}^{(+)} e^{\kappa_{2z}d} + E_{2x}^{(-)} e^{-\kappa_{2z}d} = E_{3x} e^{-\kappa_{3z}d};$$
$$H_{2y}^{(+)} e^{\kappa_{2z}d} + H_{2y}^{(-)} e^{-\kappa_{2z}d} = H_{3y} e^{-\kappa_{3z}d}.$$

By using Ampère's law:

$$\vec{\nabla} \times \vec{H} = \frac{1}{c}\frac{\partial \vec{D}}{\partial t},$$

we can relate the magnetic and electric field amplitudes which lead us to:

$$H_{1y}^{(i)} = \frac{\omega \varepsilon_1}{ck_{1z}} E_{1x}^{(i)} \,;\, H_{1y}^{(r)} = -\frac{\omega \varepsilon_1}{ck_{1z}} E_{1x}^{(r)} \,;$$

$$H_{2y}^{(+)} = \frac{i\omega \varepsilon_{xx}}{c\kappa_{2z}} E_{2x}^{(+)} \,;\, H_{2y}^{(-)} = -\frac{i\omega \varepsilon_{xx}}{c\kappa_{2z}} E_{2x}^{(-)} \,;\, H_{3y} = \frac{i\omega \varepsilon_3}{c\kappa_{3z}} E_{3x} \,.$$

Therefore, it allows us to write the following system of equations:

$$\begin{pmatrix} 1 & -1 & -1 & 0 \\ -\frac{\varepsilon_1}{k_{1z}} & -\frac{i\varepsilon_{xx}}{\kappa_{2z}} & \frac{i\varepsilon_{xx}}{\kappa_{2z}} & 0 \\ 0 & e^{\kappa_{2z}d} & e^{-\kappa_{2z}d} & -e^{-\kappa_{3z}d} \\ 0 & \frac{\varepsilon_{xx}}{\kappa_{2z}} e^{\kappa_{2z}d} & -\frac{\varepsilon_{xx}}{\kappa_{2z}} e^{-\kappa_{2z}d} & \frac{\varepsilon_3}{\kappa_{3z}} e^{-\kappa_{3z}d} \end{pmatrix} \begin{pmatrix} E_{1x}^{(r)}/E_{1x}^{(i)} \\ E_{2x}^{(+)}/E_{1x}^{(i)} \\ E_{2x}^{(-)}/E_{1x}^{(i)} \\ E_{3x}/E_{1x}^{(i)} \end{pmatrix} = \begin{pmatrix} -1 \\ -\frac{\varepsilon_1}{k_{1z}} \\ 0 \\ 0 \end{pmatrix} \Rightarrow M \begin{pmatrix} E_{1x}^{(r)}/E_{1x}^{(i)} \\ E_{2x}^{(+)}/E_{1x}^{(i)} \\ E_{2x}^{(-)}/E_{1x}^{(i)} \\ E_{3x}/E_{1x}^{(i)} \end{pmatrix} = A$$

This enables us to compute all the field components as function of the incident field $E_{1x}^{(i)}$. For the reflected field we have:

$$\frac{E_{1x}^{(r)}}{E_{1x}^{(i)}} = \frac{\det(M_1)}{\det(M)} \,;\, \frac{E_{2x}^{(+)}}{E_{1x}^{(i)}} = \frac{\det(M_2)}{\det(M)} \,;\, \frac{E_{2x}^{(-)}}{E_{1x}^{(i)}} = \frac{\det(M_3)}{\det(M)} \,;\, \frac{E_{3x}}{E_{1x}^{(i)}} = \frac{\det(M_4)}{\det(M)}$$

where $M_i$ is the matrix obtained by replacing the $i$-th column of $M$ with the vector $A$.

To fully define the electric field, we can also use Ampere's law to compare the $z$ and $x$ amplitudes. We obtain:

$$E_{1z}^{(i)} = -\frac{q}{\kappa_{1z}} E_{1x}^{(i)} \,;\, E_{1z}^{(r)} = \frac{q}{\kappa_{1z}} E_{1x}^{(r)} \,;$$

$$E_{2z}^{(+)} = -\frac{i\varepsilon_{xx} q}{\varepsilon_{zz} \kappa_{2z}} E_{2x}^{(+)} \,;\, E_{2z}^{(-)} = \frac{i\varepsilon_{xx} q}{\varepsilon_{zz} \kappa_{2z}} E_{2x}^{(-)} \,;\, E_{3z} = \frac{iq}{\kappa_{3z}} E_{3x} \,.$$

Therefore, we finally write the electric field as:

$$\vec{E}_1 = \left[ \left( E_{1,x}^{(i)}, 0, -\frac{q}{k_{1z}} E_{1x}^{(i)} \right) e^{ik_{1z}z} + \left( E_{1,x}^{(r)}, 0, \frac{q}{k_{1z}} E_{1x}^{(r)} \right) e^{-ik_{1z}z} \right] e^{i(qx-\omega t)} \,;$$

$$\vec{E}_2 = \left[ \left( E_{2,x}^{(+)}, 0, -\frac{i\varepsilon_{xx} q}{\varepsilon_{zz} \kappa_{2z}} E_{2x}^{(+)} \right) e^{\kappa_{2z}z} + \left( E_{2,x}^{(-)}, 0, \frac{i\varepsilon_{xx} q}{\varepsilon_{zz} \kappa_{2z}} E_{2x}^{(-)} \right) e^{-\kappa_{2z}z} \right] e^{i(qx-\omega t)} \,;$$

$$\vec{E}_3 = \left( E_{3,x}, 0, \frac{iq}{\kappa_{3z}} E_{3x} \right) e^{-\kappa_{3z}z} e^{i(qx-\omega t)} \,.$$

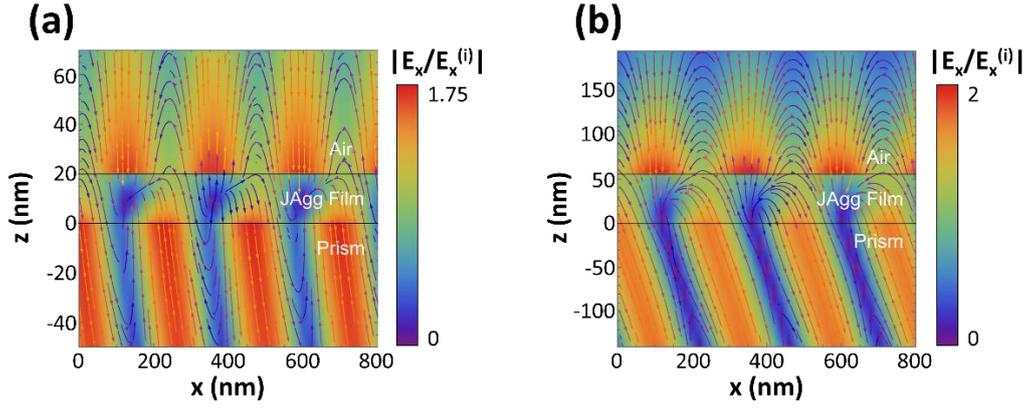

**Figure S8.** Two-dimensional colourmaps of the modulus of $E_x$ normalised by the incident electric field $E(i)$ in a cross-section of a thin film with 10 depositions with an incident angle of 55° for **(a)** j590 with $\lambda = 589$ nm and for **(b)** j620 with $\lambda = 619$ nm. Displayed arrows correspond to the electric field vector $\vec{E} = (E_x, E_z)$.

**Table S3.** Magnitude of the wavevector z-component for a j560 thin film with 15 depositions with incident angle $\theta_i = 55°$ for the wavelengths corresponding to the TMM-simulated reflectance minima of the two J-aggregate conformations.

| $\lambda$ (nm) | $k_{1z}$ (µm$^{-1}$) | $k_{2z}$ (µm$^{-1}$) | $k_{3z}$ (µm$^{-1}$) |
|---|---|---|---|
| 505 ($J^2_{560}$) | 10.83 | 6.95 - 4.85i | 9.19 |
| 540 ($J^1_{560}$) | 10.13 | 11.39 - 4.46i | 8.60 |

**Section S4. Methods**

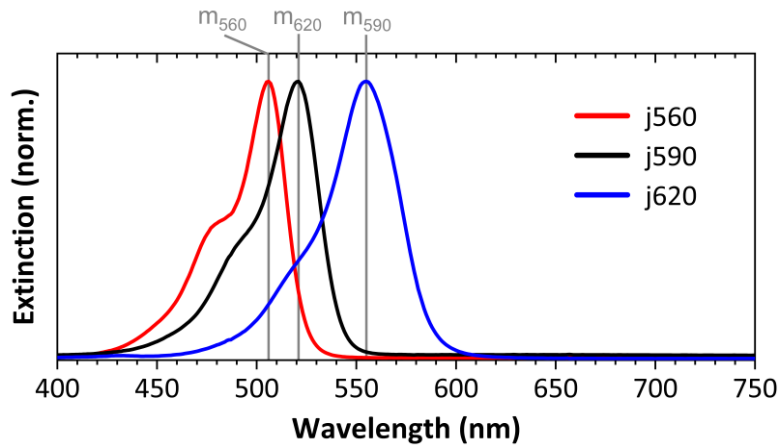

**Figure S9.** Normalized extinction spectra of different carbocyanine dyes in ethanol. Vertical lines mark the peak maxima: $m_{560}$ at 506 nm, $m_{590}$ at 521 nm, and $m_{620}$ at 555 nm. Here, m denotes the monomer peaks.

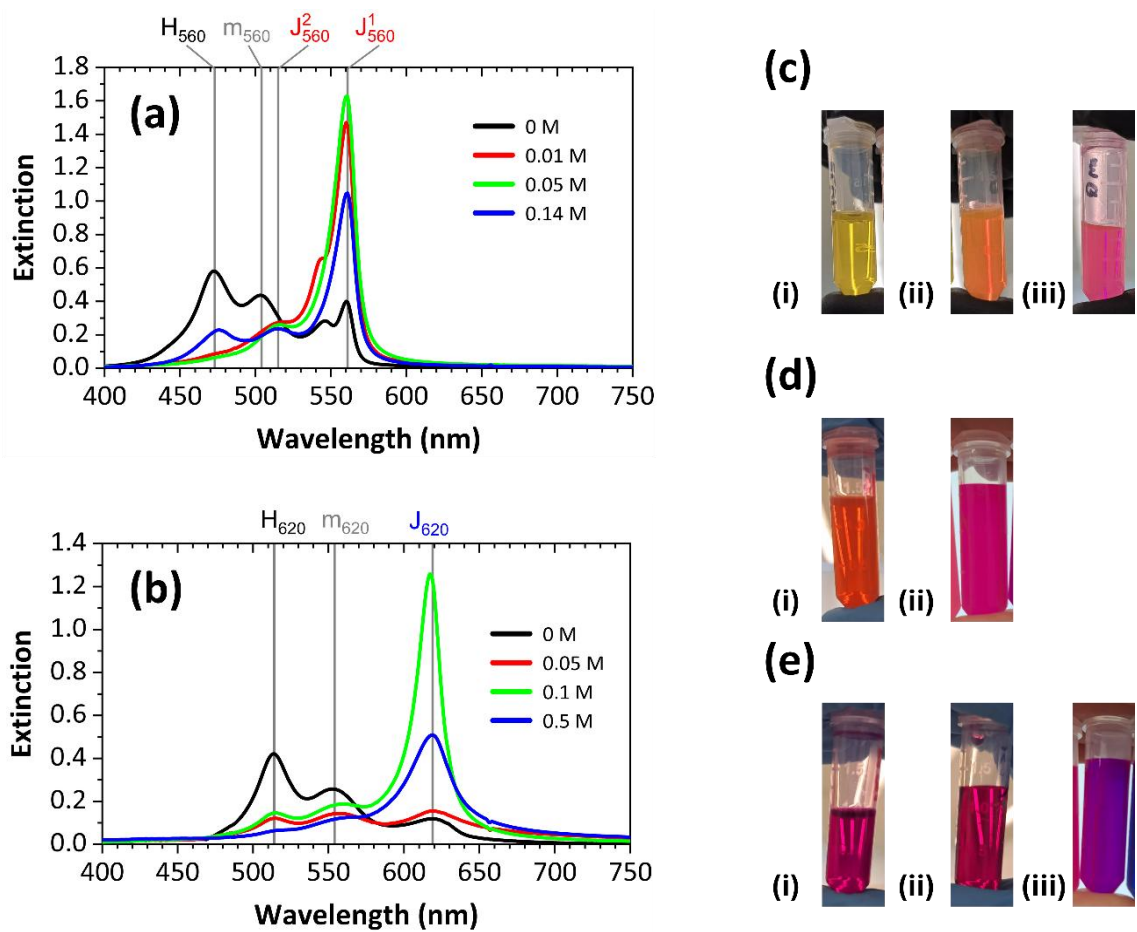

**Figure S10. (a-b)** Extinction spectra of aqueous j560 **(a)** and j620 **(b)** solutions at varying NaCl concentrations, as indicated. The monomer concentration was 100 μM. Vertical lines mark the peak maxima: $H_{560}$ at 473, $m_{560}$ at 506, $J^2_{560}$ at 515, $J^1_{560}$ at 561, $H_{620}$ at 513, $m_{620}$ at 555, and $J_{620}$ at 619 nm. Here, H denotes H-aggregate peaks, J denotes J-aggregate peak and m denotes the monomer peaks. **(c-e)** Photographs of dissolved carbocyanines: **(c)** j560, **(d)** j590, and **(e)** j620 in (i) ethanol, (ii) milli-Q water, and (iii) milli-Q water with optimal ionic strength (NaCl) to induce J-aggregate formation.